\documentclass[prd,
twocolumn,
superscriptaddress,showpacs,nofootinbib,%
tightenlines
]{revtex4}
\usepackage{graphics}
\usepackage{epsfig}
\begin{document}
\preprint{MKPH-T-04-4}
\title{Universality of the rho-meson coupling in effective field theory}
\author{D.~Djukanovic}
\affiliation{Institut f\"ur Kernphysik, Johannes
Gutenberg-Universit\"at, D-55099 Mainz, Germany}
\author{M.~R.~Schindler}
\affiliation{Institut f\"ur Kernphysik, Johannes
Gutenberg-Universit\"at, D-55099 Mainz, Germany}
\author{J.~Gegelia}
\affiliation{Institut f\"ur Kernphysik, Johannes
Gutenberg-Universit\"at, D-55099 Mainz, Germany}
\affiliation{High Energy Physics Institute,
Tbilisi State University,
University St.~9, 380086 Tbilisi, Georgia}
\author{G.~Japaridze}
\affiliation{Center for Theoretical Studies of Physical Systems,
Clark Atlanta University, Atlanta, GA 30314, USA}
\author{S.~Scherer}
\affiliation{Institut f\"ur Kernphysik, Johannes
Gutenberg-Universit\"at, D-55099 Mainz, Germany}
\begin{abstract}
   It is shown that both the universal coupling of the $\rho$ meson and
the Kawarabayashi-Suzuki-Riadzuddin-Fayyazuddin expression for the
magnitude of its coupling constant follow from the requirement that
chiral perturbation theory of pions, nucleons, and $\rho$ mesons
is a consistent effective field theory.
   The prerequisite of the derivation is that all ultraviolet divergences
can be absorbed in the redefinition of fields and the available parameters
of the most general effective Lagrangian.
\end{abstract}
\date{March 5, 2004}
\maketitle

   Already in the 1960s, vector mesons were discussed in the framework
of phenomenological low-energy chiral Lagrangians
\cite{Schwinger:1967tc,Wess:1967jq,Weinberg:de}.
   For the details of the construction of chirally invariant effective
Lagrangians describing the interaction of vector mesons with pseudoscalars
and baryons see, e.g.,
Refs.~\cite{Meissner:1987ge,Bando:1988br,Harada:2003jx,
Gasser:1984yg,Gasser:1988rb,Ecker:yg,Ecker:1988te,Borasoy:1996ds,Birse:1996hd}.
   In Ref.~\cite{Birse:1996hd}, different formulations of vector-meson
effective theories were shown to be equivalent.
   In the massive Yang-Mills approach (for a review see, e.g.,
Ref.\ \cite{Meissner:1987ge}) vector mesons are treated as gauge bosons of
local chiral symmetry (with symmetry breaking mass terms added by hand).
   This scenario implies that the $\rho$ meson couples universally,
i.e.~with the same strength, to fermions and pseudoscalars.
   In the so-called hidden chiral symmetry approach (see, e.g., Refs.\
\cite{Bando:1988br,Meissner:1987ge,Harada:2003jx}) $\rho$-coupling
universality is obtained only with a specific choice of a free parameter
of the Lagrangian.
   Although the hypotheses of dynamical bosons of both approximate and
hidden local chiral symmetry are attractive, there is, as was
emphasized in Ref.~\cite{Ecker:1988te}, no proof for the
existence of such gauge bosons of local chiral symmetry in QCD.

   Besides the construction of the effective Lagrangian, a consistent
effective field theory (EFT) program requires a systematic power
counting which allows one to organize the perturbation series.
   For example, within the extended on-mass-shell renormalization scheme of
Ref.~\cite{Fuchs:2003qc} it is possible to consistently include virtual
(axial-) vector mesons in a manifestly Lorentz-invariant formulation
of the EFT \cite{Fuchs:2003sh},
provided they appear only as internal lines in Feynman diagrams
involving soft external pions and nucleons with small three-momenta.
   Moreover, a consistent power counting also exists within the reformulated
version \cite{Schindler:2003xv} of the infrared renormalization of
Becher and Leutwyler \cite{Becher:1999he}.

   In this letter we consider the effective Lagrangian of
Ref.~\cite{Weinberg:de} describing the interaction among $\rho$
mesons, pions, and nucleons.
   In principle, the Lagrangian contains {\em all} interaction terms which
respect Lorentz invariance, the discrete symmetries, and chiral symmetry.
   As was stressed in Ref.~\cite{Weinberg:de}, the equality of the
$\rho\pi\pi$ and the $\rho NN$ coupling constants {\it does not}
follow as a consequence of the symmetries of the Lagrangian.
   Below we perform a one-loop order analysis of the nucleon and $\rho$-meson
self-energies as well as the  $\rho\rho\rho$ and $\rho NN$
vertex functions.
   In accordance with the general principles of effective field
theory \cite{Weinberg:1978kz}, we require that all ultra-violet (UV)
divergences can be absorbed into the redefinition of fields, masses,
and coupling constants, as long as one includes every one of the infinite
number of interactions allowed by symmetries \cite{Weinberg:mt}.
   The renormalization procedure imposes consistency conditions among
the (renormalized) parameters of the Lagrangian.
   In our case, both the universal coupling of the $\rho$ meson as well
as the Kawarabayashi-Suzuki-Riazuddin-Fayyazuddin (KSRF) value of the
$\rho$-meson coupling constant \cite{Kawarabayashi:1966kd,Riazuddin:sw}
turn out to be {\em consequences} of the self-consistency conditions
imposed by the EFT approach.

   We start from the chirally invariant effective Lagrangian, including
vector mesons, in the form given by
Weinberg \cite{Weinberg:de},
\begin{eqnarray}
{\cal L}&=&\frac{1}{2} \partial_\mu \pi^a_0\partial^\mu \pi^a_0
-\frac{M^{2}_0}{2}\pi^a_0 \pi^a_0\nonumber\\
&&  + \bar\Psi_0 \left( i\gamma^\mu\partial_\mu - m_0 \right)\Psi_0
\nonumber\\
&& -\frac {1}{4}F^a_{\mu\nu 0} F^{a\mu\nu}_0
+\frac {1}{2} M_{\rho 0}^2 \rho^a_{\mu 0} \rho^{a\mu}_0
\nonumber\\
&& +g_{\rho\pi\pi 0}\epsilon^{abc}\pi^a_0\partial_\mu\pi^b_0\rho^{c\mu}_0
\nonumber\\
&&+g_{\rho NN0}\bar\Psi_0\gamma^\mu \frac{\tau^a}{2}\Psi_0 \rho^a_{\mu 0}
+ {\cal L}_1,
\label{lagrangian}
\end{eqnarray}
where $\pi^a_0$ and $\rho^a_{\mu 0}$ are isospin triplets of pion and
$\rho$-meson fields with masses $M_0$ and $M_{\rho 0}$,
respectively, and $\Psi_0$ is an isospin doublet of nucleon fields with
mass $m_0$ \cite{f1}.
   The field strengths are defined as
$F^a_{\mu\nu 0}=\partial_\mu \rho^a_{\nu 0} -\partial_\nu
\rho^a_{\mu 0} +g_0 \epsilon^{abc} \rho^b_{\mu 0} \rho^c_{\nu 0}$,
where $g_{\rho NN0}\equiv g_0$ follows from chiral symmetry
\cite{Weinberg:de}.
   Finally, ${\cal L}_1$ contains an infinite number of terms
allowed by the symmetries of the theory \cite{Weinberg:1978kz,Weinberg:mt}.
   In Eq.~(\ref{lagrangian}) the subscripts 0 stand for bare quantities.
   In principle, a $\rho\rho\pi\pi$ interaction would also have a
dimensionless coupling constant, but it is not included in the Lagrangian
of Eq.\ (\ref{lagrangian}), because it is not consistent with chiral
symmetry in the present parametrization of fields \cite{Weinberg:de}.

   In order to establish relations among the renormalized coupling
constants pertaining to the Lagrangian of Eq.~(\ref{lagrangian}),
we analyze the renormalization of the coupling constant
$g_0$ using dimensional regularization in combination with the minimal
subtraction (MS) scheme (for a definition see, e.g.,
Ref.~\cite{Collins:xc}).
   However, our findings below do not depend on the choice of a specific
renormalization scheme.
   For that purpose we rewrite the Lagrangian in terms of the renormalized
fields $\pi^a$, $\Psi$, and $\rho_\mu^a$ as well as the renormalized
parameters $g$, $M$, $m$, and $M_\rho$,
\begin{eqnarray}
\pi_0^a &=& \sqrt{Z_\pi} \pi^a,\quad Z_\pi=1+\delta Z_\pi,\nonumber\\
\Psi_0 &=& \sqrt{Z_\Psi} \Psi,\quad
Z_\Psi=1+\delta Z_\Psi\, \nonumber \\
\rho_0^{a\mu} &=& \sqrt{Z_\rho}\rho^{a\mu},\quad
Z_\rho=1+\delta Z_\rho,\nonumber\\
g_0&=&g+\delta g,\nonumber\\
g_{\rho\pi\pi 0}&=&g_{\rho\pi\pi}+\delta g_{\rho\pi\pi}, \nonumber\\
M_0^2(1+\delta Z_\pi)&=&M^2+\delta M^2,\nonumber\\
m_0(1+\delta Z_\Psi)&=& m+\delta m, \nonumber\\
M_{\rho 0}^2 (1+\delta Z_\rho)&=& M_\rho^2+\delta M_\rho^2.
\label{renpar}
\end{eqnarray}
   Using Eq.\ (\ref{renpar}), we re-express the Lagrangian of
Eq.\ (\ref{lagrangian}) as \cite{Collins:xc}
\begin{displaymath}
{\cal L}={\cal L}_{\rm basic}+{\cal L}_{\rm ct}+\tilde{\cal L}_1,
\end{displaymath}
with the basic Lagrangian
\begin{eqnarray}
\label{lbasic}
{\cal L}_{\rm basic}&=&\frac{1}{2}\partial_\mu \pi^a\partial^\mu \pi^a
-\frac{M^{2}}{2}\pi^a \pi^a\nonumber\\
&&+\bar\Psi \left( i\gamma^\mu\partial_\mu - m \right)\Psi\nonumber\\
&&-\frac{1}{4}A^a_{\mu\nu} A^{a \mu\nu} +\frac {1}{2} M_{\rho}^2
\rho^a_{\mu} \rho^{a \mu}
\nonumber\\
&&+ g_{\rho\pi\pi}\epsilon^{abc}\pi^a\partial_\mu\pi^b \rho^{c\mu}\nonumber\\
&&-g\epsilon^{abc}\partial_\mu\rho_\nu^a \rho^{b\mu}\rho^{c\nu}
\nonumber\\
&&-\frac{1}{4} g^2 \epsilon^{abc}\epsilon^{ade} \rho^b_\mu
\rho^c_\nu \rho^{d \mu}\rho^{e \nu}\nonumber\\
&&+g \bar\Psi\gamma^\mu \frac{\tau^a}{2}\Psi \rho^a_{\mu},
\end{eqnarray}
the counterterm Lagrangian
\begin{eqnarray}
\lefteqn{{\cal L}_{\rm ct}=
-\frac{\delta Z_\rho}{4}\ A^a_{\mu\nu}A^{a \mu\nu}}\nonumber\\
&&+\left[\delta g_{\rho\pi\pi}
+g_{\rho\pi\pi}\left(\frac{\delta Z_\rho}{2}+\delta Z_\pi\right)\right]
\epsilon^{abc}\pi^a\partial_\mu\pi^b \rho^{c\mu}\nonumber\\
&&-\left(\delta g+\frac{3}{2} g \delta Z_\rho\right)
\epsilon^{abc}\partial_\mu\rho_\nu^a \rho^{b\mu}\rho^{c\nu}\nonumber\\
&&+\left[\delta g+g\left(\frac{\delta Z_\rho}{2}+ \delta Z_\Psi\right)\right]
\bar\Psi\gamma^\mu \frac{\tau^a}{2}\Psi \rho^a_{\mu},
\label{lct}
\end{eqnarray}
and the residual piece $\tilde{\cal L}_1$.
   In Eqs.\ (\ref{lbasic}) and (\ref{lct}), we defined
$A^a_{\mu\nu}\equiv\partial_\mu \rho^a_{\nu} -\partial_\nu
\rho^a_{\mu}$, and in ${\cal L}_{\rm ct}$ we only displayed those
counterterms explicitly which are relevant for the subsequent discussion.
   All remaining counterterms are included in $\tilde{\cal L}_1$.

\begin{figure}
\begin{center}
\epsfig{file=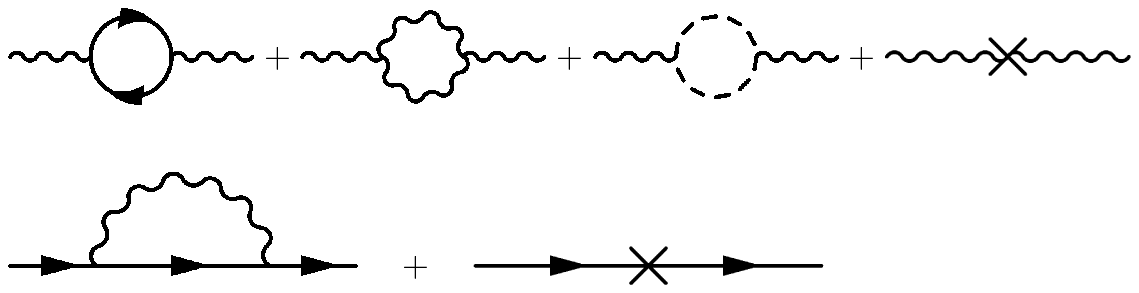,width=\linewidth}
\caption[]{\label{fse:fig} Self-energy diagrams. Solid, wiggly, and
dashed lines correspond to nucleons, $\rho$ mesons, and pions,
respectively.}
\end{center}
\end{figure}

   In the following, we will first derive the universality of the $\rho$
coupling, i.e., $g_{\rho\pi\pi}=g$.
   To that end, let us consider the $\rho\rho\rho$ and $\rho \bar\Psi\Psi$
vertex functions (amputated Green's functions) at one-loop order.
   At that order there are two types of contributions: the genuine one-loop
diagrams with basic interaction vertices and the tree-level diagrams with
one-loop order counterterms.
   We will make use of the fact that the combination of these two types of
contributions must lead to UV finite results.

\begin{figure}
\begin{center}
\epsfig{file=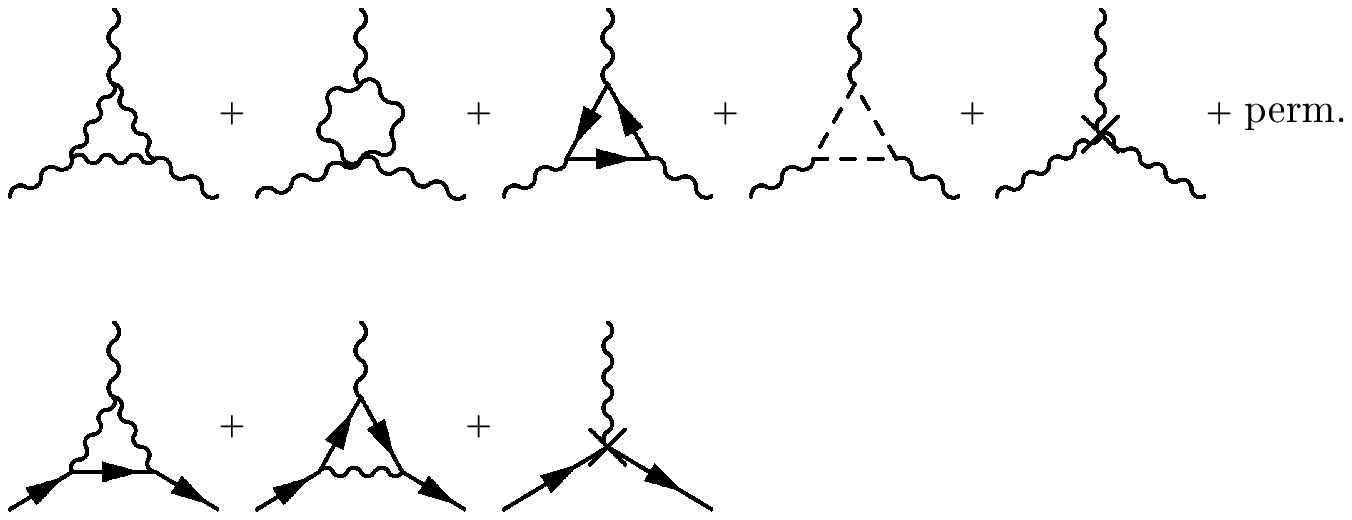,width=\linewidth}
\caption[]{\label{vertex:fig} Vertex diagrams. Solid, wiggly, and
dashed lines correspond to nucleons, $\rho$ mesons, and pions,
respectively.}
\end{center}
\end{figure}

   Up to and including one-loop order, the renormalization constant of the
$\rho$-meson field receives contributions of the form
\begin{equation}
\label{wfr}
\frac{1}{2}\delta Z_\rho=z_1 g^{2}+z_2 g_{\rho\pi\pi}^2,
\end{equation}
where the $z_1$ term corresponds to $\rho$-meson and fermion loop
contributions, and the $z_2$ term to the pion-loop contribution,
respectively (see Fig.\ \ref{fse:fig}).
   We combine the contribution of the counterterm diagram generated by
the $\rho\rho\rho$ term in the third line of Eq.~(\ref{lct}) with the
divergent parts of the one-loop contributions to the $\rho\rho\rho$ vertex
given in Fig.~\ref{vertex:fig}.
   Requiring that the result vanishes, we obtain
\begin{equation}
\label{aaacanc}
\Gamma_1 g^3+\Gamma_2 g_{\rho\pi\pi}^3-\delta g-\frac{3}{2} g\delta Z_\rho=0,
\end{equation}
where the $\Gamma_1$ term is generated by $\rho$-meson
and fermion loops, and the $\Gamma_2$ term by the pion loop, respectively.
   Substituting Eq.~(\ref{wfr}) in Eq.~(\ref{aaacanc}) and solving
for $\delta g$ we obtain
\begin{equation}
\label{dg}
\delta g=\Gamma_1 g^3 + \Gamma_2\ g_{\rho\pi\pi}^3 - 3 z_1 g^3 - 3
z_2 g g_{\rho\pi\pi}^2.
\end{equation}

   Next, we go through analogous arguments for the $\rho NN$ term.
   First, the one-loop $\rho$-meson contribution to the
renormalization constant of the $\Psi$ field, in terms of the
renormalized couplings, reads
\begin{equation}
\label{wfrf}
\frac{1}{2}\delta Z_\Psi=w_1 g^{2}.
\end{equation}
   Again, we combine the divergent parts of the one-loop contributions to the
$\rho\bar\Psi\Psi$ vertex given in Fig.~\ref{vertex:fig} with the
contribution of the counterterm diagram generated by the
$\rho\bar\Psi\Psi$ term in the last line of
Eq.~(\ref{lct}) and require the result to vanish:
\begin{equation}
D_1 g^3+\delta g +g \frac{1}{2}\delta Z_\rho+ g \delta Z_\Psi=0.
\label{aaacancf}
\end{equation}
Substituting Eq.~(\ref{wfrf}) in Eq.~(\ref{aaacancf}) and solving
for $\delta g$ we obtain
\begin{equation}
\delta g=-D_1 g^3 -\left( z_1+2 w_1\right) g^3 - z_2 \ g\
g_{\rho\pi\pi}^2. \label{dgf}
\end{equation}
   In order to have a self-consistent theory, the two expressions for
$\delta g$ given by Eqs.~(\ref{dg}) and (\ref{dgf}) must coincide.
   Calculating $z_1$, $z_2$, $w_1$, $\Gamma_1$, $\Gamma_2$, and $D_1$
explicitly \cite{Djukanovic} and comparing Eqs.~(\ref{dg}) and (\ref{dgf}) we
obtain:
\begin{equation}
\label{couplrelo}
g_{\rho\pi\pi}^3=g g_{\rho\pi\pi}^2.
\end{equation}
   Equation (\ref{couplrelo}) has a trivial solution $g_{\rho\pi\pi}=0$, which
corresponds to the EFT without pions, and the non-trivial solution
\begin{equation}
g_{\rho\pi\pi}=g.
\label{couplrel}
\end{equation}
   In other words, from the EFT standpoint the universality of the $\rho$
coupling, $g_{\rho\pi\pi}=g$, is a {\em consequence} of the consistency
conditions imposed by the requirement of perturbative
renormalizability \cite{f2}.
   From this point of view, universality neither appears to be something
which has to be postulated \cite{Kawarabayashi:1966kd}, nor is it obtained
as the result of a dynamical principle such as vector-meson dominance
\cite{Riazuddin:sw}.

    As has been pointed out in, e.g., Refs.~\cite{Weinberg:de,Ecker:yg},
chiral symmetry specifies the combination of the vector-meson mass term
and the $\rho \pi\pi$ coupling that should appear in the
Lagrangian.
    This combination reads
\begin{equation}
\frac{M^2_0}{2} \rho_{\mu 0}^a \rho^{a \mu}_0-\frac{1}{2} M_0^2
g_0^{-1}F_{0}^{-2} \epsilon^{abc}\pi^a_0\partial_\mu\pi^b_0
\rho^{c \mu}_0,
\label{rhopipi}
\end{equation}
where $F_{0}$ denotes the bare pion-decay constant in the chiral limit.
   Comparing Eq.~(\ref{rhopipi}) with the Lagrangian of
Eq.~(\ref{lagrangian}) it follows that, in order to retain chiral
symmetry, the relation
\begin{equation}
g_{\rho\pi\pi 0}=\frac{M_{\rho 0}^2}  {2 g_0 F_{0}^2}
\label{ksrf0}
\end{equation}
should hold.
   Taking Eq.~(\ref{couplrel}) into account and performing a loop expansion
of bare quantities, $g_0=g+O(\hbar)$ etc.,
we obtain from Eq.~(\ref{ksrf0}), in terms of renormalized quantities,
\begin{equation}
g^2=\frac{M_{\rho }^2}  {2 F^2}. \label{ksrf}
\end{equation}
   Equation (\ref{ksrf}) is the well-known
Kawarabayashi-Suzuki-Riazuddin-Fayyazuddin relation
\cite{Kawarabayashi:1966kd,Riazuddin:sw}
which was originally derived by combining current algebra with soft-pion
techniques and either assuming universality \cite{Kawarabayashi:1966kd} or
assuming vector-meson dominance in conjunction with the conserved vector
current hypothesis \cite{Riazuddin:sw}.
   In the EFT framework, this relation is a natural consequence of
chiral symmetry and the consistency conditions which are imposed
on the parameters of the effective Lagrangian by the requirement
of renormalizability.

We would like to emphasize that our analysis of the
renormalizability of the considered EFT is only partial (the
complete analysis is beyond the scope of this letter). For
example, in addition to the divergences explicitly analyzed here,
the considered diagrams contain divergences requiring counterterms
which have a non-renormalizable structure (in the traditional
sense). For the consistent EFT it should be possible to absorb all
these divergences in the redefinition of the parameters of ${\cal
L}_1$. Furthermore, the applied dimensional regularization
implicitly subtracts all power-law divergences. In a complete
analysis it would be necessary to show that these divergences can
be absorbed in the redefinition of the parameters and fields of
the most general effective Lagrangian. Although we believe that,
in the sense of renormalizability, there exists a consistent EFT
of nucleons, pions and vector-mesons, to the best of our knowledge
this never has been proven. Therefore, strictly speaking, our
results should be interpreted that, given the existence of a
consistent EFT, universality and the KSRF relation necessarily
hold.

   To summarize, we have considered the most general effective Lagrangian for
the interaction of $\rho$ mesons with pions and nucleons, respecting Lorentz
invariance, the discrete symmetries, and chiral symmetry.
   While usually the universal $\rho$ coupling is taken as an
additional {\em assumption}, in the framework of EFT it
has a deep foundation, namely the consistency of EFT {with respect to
renormalization}.
   In addition, combining chiral symmetry and the consistency of EFT
naturally generates the KSRF relation among the renormalized $\rho$-coupling
constant, $\rho$-meson mass, and pion-decay constant.

\begin{acknowledgments}
The work of J.~Gegelia was supported by the
Deutsche Forschungsgemeinschaft (SFB 443).
G.~Japaridze thanks the National Science Foundation (NSF-G067771) for support.
\end{acknowledgments}


\begin{thebibliography}{100}
\bibitem{f1}
We only analyze interaction terms with
dimensionless coupling constants.
   The explicit inclusion of non-renormalizable interaction terms is not
necessary for the purpose of this work.
   Note that, although the perturbation series is organized as a systematic
expansion in small momenta and the pion mass,
the renormalization program still follows the standard procedure of absorbing
divergences in field redefinitions and the renormalization of masses and
coupling constants.
\bibitem{f2}
Let us stress that we have derived Eq.\
(\ref{couplrel}) to leading order, i.e., we have shown that
$g_{\rho\pi\pi}=g+O(\hbar)$.
\end{thebibliography}
\end{document}